\documentclass[showpacs,twocolumn,prl,superscriptaddress]{revtex4}
\usepackage{ulem}
\usepackage{graphicx}
\usepackage{amssymb}
\usepackage{amsmath}
\usepackage{amsfonts}
\usepackage{mathrsfs}
\renewcommand{\epsilon}{\varepsilon}
\newcommand{\figurewidth}{0.43\textwidth}

\begin{document}
\title{Scaling exponents of Forced Polymer Translocation through a nano-pore}
\author{Aniket Bhattacharya}
\altaffiliation[]{
Author to whom the correspondence should be addressed}
\email{aniket@physics.ucf.edu}
\affiliation{Department of Physics, University of Central Florida, Orlando, Florida
32816-2385, USA}

\author{William H. Morrison}
\affiliation{Department of Physics, University of Central Florida, Orlando, Florida
32816-2385, USA}

\author{Kaifu Luo}
\affiliation{Department of Applied Physics, Helsinki University of Technology,
P.O. Box 1100, FIN-02015 TKK, Espoo, Finland}

\author{Tapio Ala-Nissila}
\affiliation{Department of Applied Physics, Helsinki University of Technology,
P.O. Box 1100, FIN-02015 TKK, Espoo, Finland}
\affiliation{Department of Physics, Box 1843, Brown University, Providence,
Rhode Island 02912-1843, USA}
\author{See-Chen Ying}
\affiliation{Department of Physics, Box 1843, Brown University, Providence, Rhode Island
02912-1843, USA}
\author{Andrey Milchev}
\affiliation{Institute of Physical Chemistry, Bulgarian Academy of Sciences,
Georgi Bonchev Street, Block 11, 1113 Sofia, Bulgaria}
\author{Kurt Binder}
\affiliation{Institut f\"ur Physik, Johannes Gutenberg Universit\"at Mainz, Staudinger Weg 7, 
55099, Mainz, Germany}
\date{\today}
\begin{abstract}
We investigate several scaling properties of a translocating homopolymer through a thin pore driven 
by an external field present inside the pore only using Langevin Dynamics (LD) simulations in 
three dimensions (3D). Motivated by several recent theoretical and numerical studies 
that are apparently at odds with each other, we determine the chain length ($N$) dependence 
scaling exponents of the average translocation time $\langle \tau \rangle$, 
the average velocity of the center of mass $\langle v_{CM} \rangle $, and the effective 
radius of gyration $\langle \tilde {R}_g \rangle $ during the translocation process defined as 
$\langle \tau \rangle \sim N^\alpha $, $\langle v_{CM} \rangle \sim N^{-\delta}$, and 
$\tilde {R}_g \sim N^{\bar{\nu}}$ respectively, and the scaling exponent of 
the translocation coordinate ($s$-coordinate) as a function of the translocation time  
$\langle s^2(\tau) \rangle \sim \tau^{\beta}$. We find $\alpha = 1.36 \pm 0.01$, $\beta = 1.60 \pm 0.01$
for $\langle s^2(\tau) \sim \tau^\beta \rangle$ and  $\bar\beta = 1.44 \pm 0.02$ 
for $\langle \Delta s^2(\tau)\rangle \sim \tau^{\bar\beta} $,  
$\delta = 0.81 \pm 0.04$, and $\bar{\nu} \simeq \nu=0.59 \pm 0.01$, where $\nu$ is the equilibrium Flory 
exponent in 3D. Therefore, we find that $ \langle \tau  \rangle \sim N^{1.36}$ is consistent with  
the estimate of 
$\langle \tau \rangle \sim \langle R_g \rangle / \langle v_{CM} \rangle$. However, as observed 
previously in MC calculations by Kantor and Kardar (Y. Kantor and M. Kardar, 
Phys. Rev. E, {\bf 69}, 021806 (2004)) we also 
find the exponent $\alpha = 1.36 \pm 0.01 < 1+\nu $. We also observe that 
$\alpha = 1.36$ is in closer agreement with  $\alpha \simeq (1+2\nu)/(1+\nu)$ as 
recently proposed  by 
Vocks \textit{et al.} (H. Vocks, D. Panja, G. T. Barkema, and R. C. Ball, 
\textit{J. Phys.: Condens. Matter} {\bf 20}, 095224 (2008)).  
We also discuss the dependence of the scaling exponents on the pore geometry for the range of 
$N$ studied here.
\end{abstract}
\maketitle
\section{INTRODUCTION}
Translocation of biopolymers accros a biomembrane, \textit{ e.g.},
transport of RNA molecules out of a nucleus, invasion of viruses into cells, \textit{ etc.}, 
are ubiquitous and important processes in biological systems\cite{Alberts}. 
Recently voltage driven translocation 
of a single stranded DNA through a $\alpha$-hemolysin pore in 
biomembrane\cite{Kasianowitch96}, and subsequently 
double stranded DNA through synthetic silicon nanopores\cite{Li01}
have stimulated a lot of activities as the phenomenon is rich 
in fundamental science involved and  its prospective technical applications for detecting 
DNA/RNA sequences.  
While it is the attributes of heteropolymer translocation that are the key ingredients for prospective 
new sequencing methods, these experiments have generated stimulating theoretical and 
numerical studies directed towards fundamental physics of homopolymer translocation 
through a nanopore. An important question that has been repeatedly raised is how does the 
average translocation time scale with the 
chain length and what is the equation of motion that describes the situation adequately. 
Approaches using Fokker-Planck equation with entropic barrier term incorporated in the 
free energy have generated useful insights to the 
problem\cite{Sung96}-\cite{Wolterink06}
More recently partial Fokker-Planck equation (PFPE) 
has been suggested is the natural language of the problem\cite{Dubbledam1,Dubbledam2}.  
Quite naturally, a number of simulational studies have 
been directed to test predictions of these theories\cite{Chuang01}-\cite{Kaifu0805}. \par

This paper is aimed at determining the relevant scaling exponents of forced translocation 
of a homopolymer through a nanopore by carrying out large scale Langevin dynamics (LD) simulations 
in three dimensions (3D) and comparing the findings with those predicted by theoretical arguments. 
We look at the arguments for the unbiased case first as it serves as the 
reference for extending the theoretical arguments in the presence of an external field. 
Naturally, the equilibrium radius of gyration $R_g \sim N^\nu$ of a chain of length $N$, 
where $\nu$ is the Flory exponent, is used as the relevant length scale in all the theories.
The first theoretical argument came from Chuang 
\textit{et al.} \cite{Chuang01} who predicted that for the unbiased translocation 
the mean translocation time should  scale 
in the same manner as a freely diffusing chain so that  
$\langle \tau \rangle \sim R_g^2/D \sim N^{1+2\nu}$, 
assuming the diffusion coefficient $D \sim 1/N$ appropriate for the 
free-draining limit (no hydrodynamic interaction).  
In this theory it is argued that the Rouse relaxation serves as the lower bound and 
in presence of a nanopore a smaller amplitude accounts for 
the slowness of the process\cite{Milchev04}. 
This theory also predicts that the scaling exponent of the reaction coordinate  
defined as $\langle s^2(\tau) \rangle \sim \tau^\beta$ is given by $\beta = 2/\alpha$.
As usual\cite{Sung96}-\cite{Wolterink06} we denote by $s(\tau)$ 
the monomer that is inside the pore at time $\tau$.    
Noticeably, the theory is essentially very simple and the exponents are functions of 
$\nu$ only with $\alpha = 1+2\nu$, $\beta = 2/(1+2\nu)$ so that $\alpha\beta = 2$. 
In two dimensions $(2D)$ $\nu = 0.75$ leads to   
$\alpha = 2.5$ and $\beta = 0.8$ respectively. In three dimensions (3D) $\nu=0.588$ leads to  
$\alpha = 2.2 $ and $\beta = 0.92$ respectively.  
The theory put forward by Dubbledam 
\textit{et al.} invokes an additional surface exponent term $\gamma_1$\cite{gamma} 
so that for the diffusive case this theory predicts\cite{Dubbledam1,Dubbledam2}
$\alpha = 2(1+\nu) - \gamma_1$ and $\beta =2/\alpha $. 
For unbiased translocation this theory also predicts the product $\alpha\beta = 2$. 
Several recent numerical studies in $2D$
\cite{Chuang01,Kaifu06a,Kaifu06b,Huopaniemi06a} and one in 3D\cite{Liao} supports 
Chuang \textit{et al.}, while Dynamic Monte Carlo (DMC) results by Dubbledam \textit{et al.} 
report $\alpha = 2.5$ and $\beta = 0.8$ in 3D  
which contradicts Chuang \textit{et al.} and supports their own prediction\cite{Dubbledam1}. 
While all the simulation studies verify $\alpha\beta = 2.0$, 
recent theories by Panja \textit{et al.} and Vocks \textit{et al.} pointed out the role of  
decay time of monomer density near the pore and argues that the translocation time 
is anomalous up to the Rouse time $t_R \sim N^{1+2\nu}$, and becomes diffusive 
afterwards\cite{Panja072,Vocks}. Therefore, for the unbiased translocation 
the collective numerical results do not support any of the proposed theories completely.\par

Let us now look at the theoretical studies of driven translocation whose numerical verification including 
the underlying assumptions is the main focus of the paper.  According to  Kantor and Kardar  
$\langle \tau \rangle  \sim \langle R_g \rangle / \langle v_{CM} \rangle \sim N^{1+\nu}$, 
assuming $v_{CM} \sim 1/N$. Kantor and Kardar\cite{Kantor04} 
argued that since the chain is only driven 
at one point inside the narrow pore, the accompanying change in its shape due to the bias 
is insignificant for the rest of the chain and therefore, the chain in this case is also 
described by the equilibrium Flory exponent $\nu$.  
To verify their scaling argument Kantor and Kardar carried out 
Lattice MC simulation of self-avoiding chains in $2D$ and noticed that the numerical exponent
$ \simeq 1.5 < 1+\nu = 1.75 $. They argued that finite size effects are severe in this 
case and the relation 
$\langle \tau \rangle \sim N^{1+\nu}$ should be taken as an upper bound that will be seen only for the 
extremely large chains. Vocks \textit{et al.} on the contrary, 
using arguments about memory effects in the monomer dynamics 
came up with an alternate estimate\cite{Vocks}  
$\langle \tau \rangle  \sim N^{\frac{1+2\nu}{1+\nu}}$. This seems to be consistent 
with most of the numerical data in 3D. However this estimate fail to capture 
the recent 2D simulation results using Langevin dynamics and MC simulations
\cite{Kaifu06b,Huopaniemi06a} where 
one sees a crossover of the $\alpha$-exponent from 1.5 to 1.7 (as opposed to 1.428).  
Dubbledam \textit{et al.} have extended their PFPE based theory 
for the driven translocation\cite{Dubbledam2} and came up with the following relations 
$\alpha = 2\nu +1 - \gamma_1$ and $\beta = 4/(2(1+\nu) - \gamma_1)$. 
The prediction of Dubbledam \textit{et al.} for   
the exponents are  $\alpha = 1.55$ and $\beta = 1.56$ in $2D$ and $\alpha =  1.5$, and 
and $\beta = 1.6$ in 3D respectively. The DMC results of Dubbledam \textit{et al.}  are consistent 
with this theory. However, more recent numerical results using LD and MD\cite{Kaifu0805,Aniket} 
produce similar results which are only in partial agreement with these theories.
\par  
In this paper not only we calculate these scaling exponents $\alpha$ and $\beta$ 
for the driven chain but provide insights how the scaling aspects are affected by boundary and geometric 
factors by monitoring some of the relevant time dependent quantities 
during the translocation process. This allows us to check   
how well some of the assumptions are satisfied for the driven translocation and discuss possible 
scenarios for the disagreements between the theoretical predictions and numerical studies. 
Thus far these issues have not been adequately addressed in the literature. 
\section{THE MODEL}
We have used the ``Kremer-Grest'' bead spring model to mimic a strand of 
DNA \cite{Grest1}. 
Excluded volume interaction between monomers is
modeled by a short range repulsive LJ potential
\begin{eqnarray*}
U_{LJ}(r)&=&4\epsilon [{(\frac{\sigma}{r})}^{12}-{(\frac{\sigma}
{r})}^6]+\epsilon \;\mathrm{for~~} r\le 2^{1/6}\sigma \\
        &=& 0 \;\mathrm{for~~} r >  2^{1/6}\sigma\;.
\end{eqnarray*}
Here, $\sigma$ is the diameter of a monomer, and
$\epsilon$ is the depth of the potential. The connectivity between
neighboring monomers is modeled as a Finite Extension Nonlinear
Elastic (FENE) spring with 
\[ U_{FENE}(r)=-\frac{1}{2}kR_0^2\ln(1-r^2/R_0^2)\;,\] 
where $r$ is the distance
between consecutive monomers, $k$ is the spring constant and $R_0$
is the maximum allowed separation between connected monomers.
We use the Langevin dynamics with the equation of motion 
\[ \ddot { \vec{r}} _i = - \, \vec
{\nabla} U _ i - \Gamma \dot { \vec{r}} _i \, + \vec {W} _ i (t) \;.\] 
Here $\Gamma$ is the monomer friction coefficient and
$\vec{W} _ i (t)$, is a Gaussian white noise with zero mean at a temperature T, and
satisfies the fluctuation-dissipation relation:  
\[ < \, \vec{W} _ i (t)
\cdot \vec{W} _ j (t') \, > = 6k_BT \Gamma \, \delta _{ij} \, \delta (t
- t ')\;.\]

The purely repulsive wall consists of one monolayer of LJ particles of diameter 1.5$\sigma$ on a \textit{ triangular lattice} 
at the $xy$ plane at $z=0$. The pore is created by removing the particle at the center. 
Inside the pore, the polymer beads 
experience a constant force $F$ and a repulsive potential from the inside wall of the pore. 
The reduced units of length, time and temperature are chosen to be  $\sigma$, 
$\sigma\sqrt{\frac{m}{\epsilon}}$, and $\epsilon/k_B$ respectively.  
For the spring potential we have chosen $k=30$ and $R_{ij}=1.5\sigma$, the friction co-efficient 
$\Gamma = 1.0$, and  
the temperature is kept at $1.5/k_B$ throughout the simulation.\par
  
We carried out simulations for chain lengths $N$ from $8 - 256$ for 
two choices of the biasing force $F = 4$  and $6$, respectively.  
Initially the first monomer of the chain is placed at the entry of the pore.  
Keeping the first monomer in its original position the rest of the 
chain is then equilibrated for times at least an amount proportional to the $N^{1+2\nu}$. 
The chain is then allowed to move through the pore driven by the 
field present inside the pore. When the last monomer exits the pore we stop the 
simulation and note the translocation time and 
then repeat the same for $5000$ such trials. \par

\section{SIMULATION RESULTS AND THEIR INTERPRETATION}

\begin{figure}[tb]
\begin{center}
\includegraphics[width=\figurewidth]{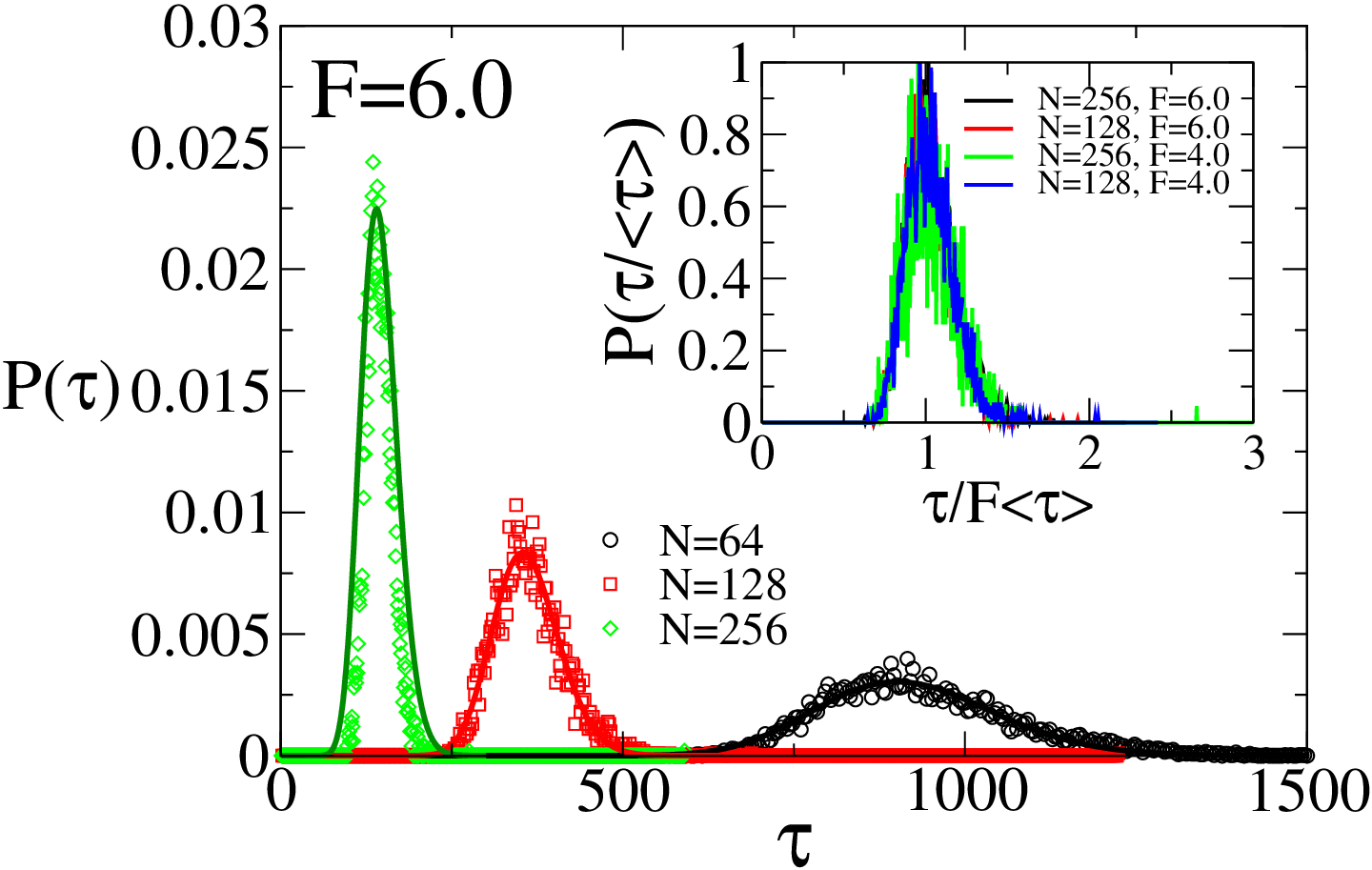}
\caption{Histogram $P(\tau)$ of flight times for chain lengths $N=$ 
64, 128, and 256 for bias F=6.0. The symbols represent simulation data and the 
solid lines are fits with a form $P(\tau) = A \tau ^{\lambda} \exp(-\mu\tau)$. 
The inset shows the corresponding scaled plots where   
the $\tau$-axis has been scaled by $F\langle \tau \rangle$ and 
the $y$-axis has been scaled by the maximum value of the histogram. }
\label{histo}       
\end{center}
\end{figure}

Typical histograms for the passage time are shown in Fig.~\ref{histo} for $F=6.0$. 
When the time axis is scaled by the mean translocation time multiplied by the 
bias $F$ and the peak of the distribution is normalized to unity, we observe (inset) a nice 
scaling of all the histograms on a single master curve. We also note that an excellent 
fit (solid lines) could be made with an expression $P(\tau) = A \tau ^{\lambda} \exp(-\mu\tau)$
for all the plots with the peak position being given by $\tau_{max} = \lambda/\mu$. 
We calculated the average translocation time from the weighted mean
$\langle \tau \rangle = \int_0^ {t_{max}}\tau  P(\tau)d \tau $, 
where $t_{max}$ for each distribution is chosen such that at $t_{max}$ the
distribution $P(\tau)$ is about 0.01 \% of its peak value. We have checked that
$\langle \tau \rangle$ calculated from the area is marginally greater than $\tau_{peak}$
obtained from $P(\tau)$. \par
\begin{figure}[tb]                
\begin{center}
\includegraphics[width=\figurewidth]{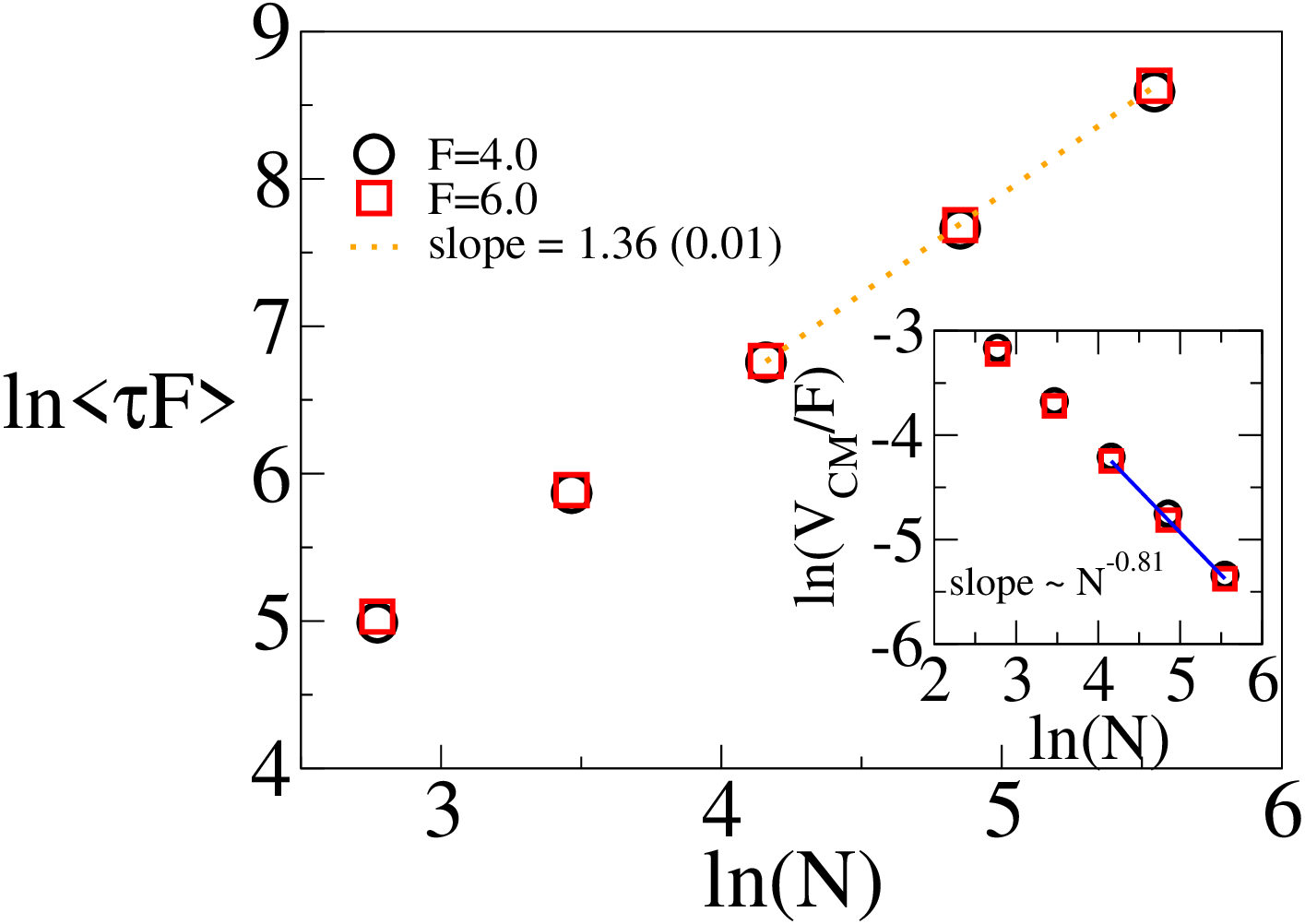}
\caption{
Scaling of the mean translocation time $\langle \tau \rangle $ (logarithmic scale) 
scaled by the applied bias $F$ as a function of chain length $N$ (logarithmic scale). 
The open  circles and squares refer to $F=4.0$ and $F=6.0$ respectively. 
The inset shows the corresponding scaling of $v_{CM}/F$. }
\label{tau}       
\end{center}
\end{figure}
\begin{figure}[tb]                
\begin{center}
\includegraphics[width=\figurewidth]{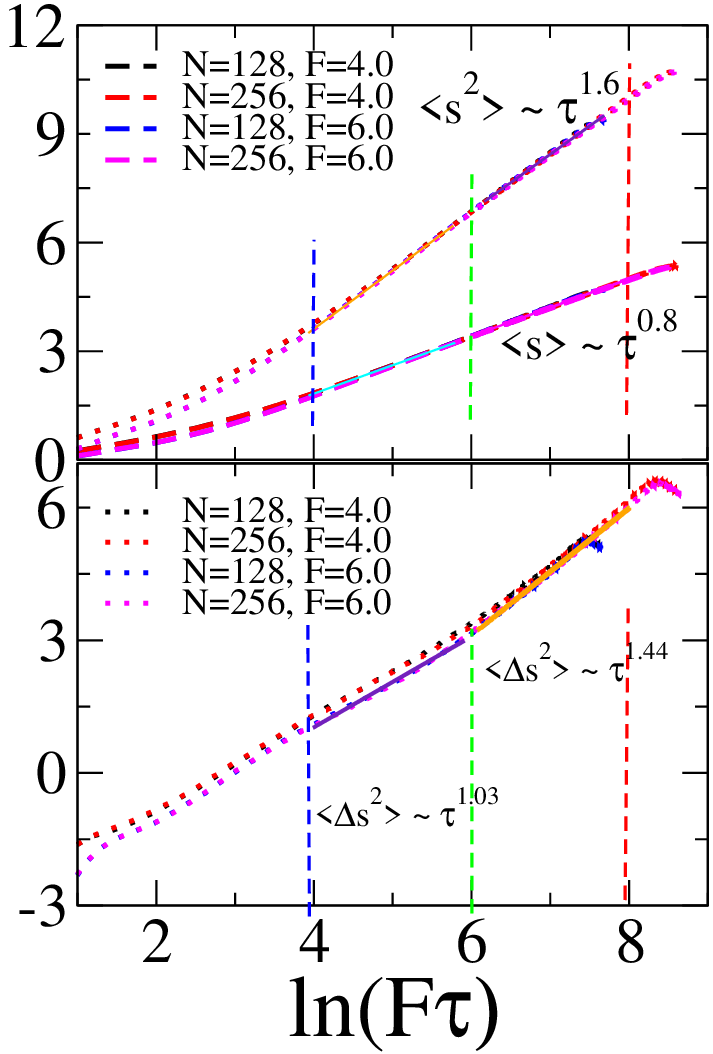}
\caption{variation of 
$\langle s^2(\tau) \rangle$ (top, dotted) and $\langle s(\tau) \rangle$ (top, dashed-dot), 
and $\langle s^2(\tau) - \langle s \rangle ^2 \rangle$ (bottom)  
as a function of the scaled translocation time $F\tau$. The black and blue colors correspond 
to chain length N=128 for F=4.0 and F=6.0 respectively. 
The red and magenta colors correspond 
to the chain length N=256 for F=4.0 and F=6.0 respectively. }
\label{anomal}
\end{center}
\end{figure}

The scaling exponent $\alpha$ of the mean translocation 
time $\langle \tau \rangle \sim N^\alpha $ is extracted by plotting the  
$\langle \tau \rangle$ as a function of $N$ shown 
in Fig.~\ref{tau}. Evidently, we find that $ \langle \tau \rangle \sim 1/F$ and 
$\langle \tau \rangle \sim N^{1.36}$. The inset of Fig.~\ref{tau} shows that the 
velocity of the center of mass increases linearly with the bias and scales as 
$v_{CM} \sim 1/N^{0.81}$. We note that $v_{CM}$ does not scale as $1/N$. 
It has been suggested that this exponent is not universal and 
depends on the width and the 
geometry of the pore\cite{Kaifu0805}. We will come back to this issue later.
The scaling exponent $\beta$ of the $s$ coordinate is shown in Fig.~\ref{anomal}. 
For clarity, we have shown results for the two largest chain lengths $N$ = 128 and 256. When 
we calculate the first and the second moments of $s(\tau)$ we find that 
$\langle s(\tau) \rangle \sim \tau^{0.8}$ and $\langle s^2(\tau) \rangle \sim \tau^{1.6}$ for a wide 
range of the translocation time (the slope remains 
the same between the blue and the green vertical windows and 
between the green and the red vertical windows respectively in Fig. 3).  
The data as a function of 
the scaled translocation time $F\tau$ show excellent collapse. 
Since $\langle s^2(\tau) \rangle \sim (\langle s(\tau) \rangle)^2 $, 
one expects to see   
$\langle \Delta  s^2(\tau) \rangle = \langle s^2(\tau) - \langle s(\tau) \rangle ^2 \rangle \sim \tau^{1.6}$
during the same time window. However, 
$\langle s^2(\tau) - \langle s(\tau) \rangle ^2 \rangle $  
reveals additional features where the slope changes from 
$\langle s^2(\tau) - \langle s(\tau) \rangle ^2 \rangle \sim \tau^{1.03} $ 
(between blue and green dashed vertical lines) to 
$\langle s^2(\tau) - \langle s(\tau) \rangle ^2 \rangle \sim \tau^{1.44}$
(between green and red vertical lines). For the forced translocation $\langle s(\tau) \rangle \ne 0$ and 
it is likely that a tiny difference of time dependence 
of the first and 2nd moment during the translocation of the chain that is not visible in 
the plot of 1st or the 2nd moment of the $s$-coordinate becomes noticeable in its fluctuation. 
Therefore, if we use the fluctuations in $s$ to define 
$\langle (s(\tau) - \langle s(\tau) \rangle)^2 \rangle \sim \tau^{\bar{\beta}} $, then from the late time slope 
(Fig. 3) then we get $\bar{\beta} = 1.44$. \par
We now compare these results with the theoretical predictions and other existing 
numerical results. The translocation exponent in 3D, according to Kantor and Kardar\cite{Kantor04}, is 
$1+\nu = 1.588$, and according to Dubbledam\cite{Dubbledam2} is $1.50$. First of all, 
as observed in $2D$ MC simulations 
by Kantor and Kardar\cite{Kantor04} we also obtain a smaller value 
of $\alpha = 1.36 \pm 0.01 < 1+\nu = 1.588 $ in 3D. Kantor and Kardar argued that a lower value 
is obtained due to finite size effects and expect that for very large chains one would find 
$1+\nu \simeq 1.59$. This bound has recently been criticized by Vocks \textit{et al.}\cite{Vocks} who 
using arguments about memory effects in the monomer dynamics  
came up with an alternate exponent estimate   
given by $\frac{1+2\nu}{1+\nu} = 1.37$. Evidently, our result is in  
agreement with this prediction. As for the exponent $\beta$ we find 
$\langle s^2(\tau)\rangle \sim \tau^{1.6}$, 
and $\langle \Delta s^2(\tau)\rangle = \langle s^2(\tau) - \langle s \rangle ^2 \rangle \sim \tau^{1.44}$ 
(if we use the later window). Therefore, with Dubbledam \textit{et al.} we  do not agree 
with the calculated value of 
$\alpha$, but Dubbledam \textit{et al.} also used $\langle s^2(\tau) \sim \tau^\beta$, 
to define the exponent 
$\beta$ and the numerical value $\beta = 1.6$ is exactly the 
same as found here. It is noteworthy that the fluctuation $\langle \Delta s^2(\tau) \rangle $ is time dependent 
and the slope of $\langle \Delta s^2(\tau) \rangle \sim \tau^{1.03}$  at early time crosses over to 
$\langle \Delta s^2(\tau) \rangle \sim t^{1.44}$ at a later time, while the slope for 
$\langle s^2(\tau) \rangle \sim \tau^{1.6}$ is constant for a wider range. If we use $\beta = 1.44$, obtained 
from the definition of fluctuation of the $s$ coordinate, then
we find the relation $\alpha\beta = 2.0$ is satisfied for the forced translocation as well. This 
trend is qualitatively the same for the simulation using a square pore\cite{Aniket}, where 
we find that $\langle \tau \rangle \sim N^{1.41}$,   
$\langle s^2(\tau)\rangle \sim \tau^{1.52}$, and $\langle \Delta s^2(\tau)\rangle \sim \tau^{1.45} $
(so that $\alpha\beta \simeq 2.0$, same as reported here if we extract $\beta$ from 
the slope of the plot $\langle \Delta s^2(\tau)\rangle \sim \tau $). 
Our results may be relevant in the context of a recent recent article by Chatelain, Kantor, and 
Kardar\cite{Chatelain} who showed that the variance of the probability distribution $P(s,t)$ 
grows subdiffusievly. \par

We now look more closely at the factors responsible for the translocation process. The expression   
$\tau \sim \langle R_g \rangle / \langle v_{CM} \rangle \sim N^{1+\nu}$   
has two components: the dependence of $v_{CM}$ on $N$ and $R_g$ on $N$ respectively. We now look at 
these two components separately.  During the driven translocation the chain does not find enough time 
to relax. Therefore, it is important to know how does
the shape of the chain vary as a function of time and how different it is compared to its equilibrium 
configuration. During the forced translocation at any instant of time only 
one segment of the entire chain feels the bias. Kantor and Kardar\cite{Kantor04} argued that 
the shape of the chain is hardly 
affected by it so that it will still be described by the equilibrium Flory 
exponent $\nu$. This argument will not be strictly valid for the model used here as 
the beads are connected by elastic bonds and it is expected that quite a few 
neighbors on either side of the 
driven bead inside the pore will be indirectly affected by it.      
     
\begin{figure}[tb]
\begin{center}               
\includegraphics[width=\figurewidth]{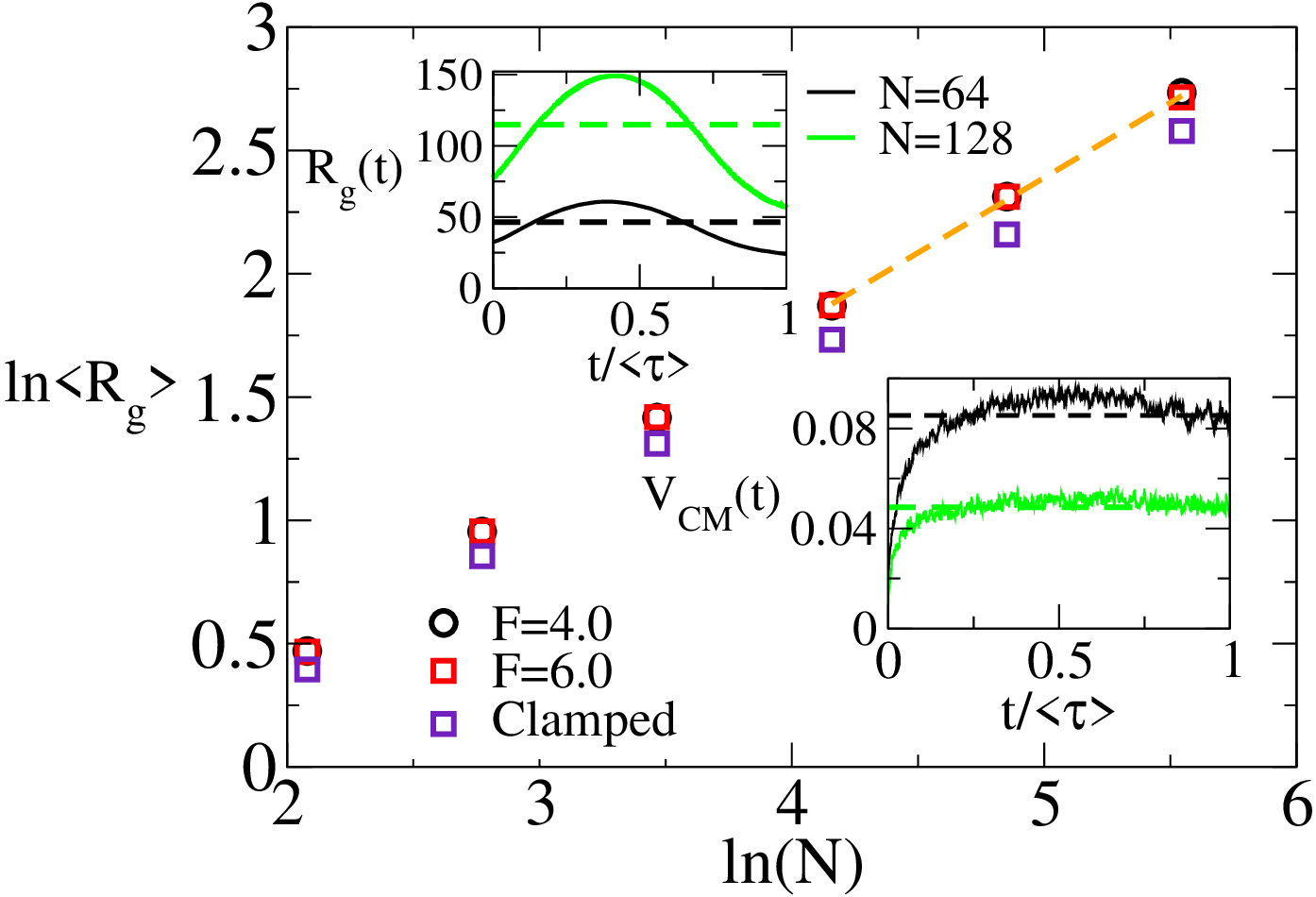}
\caption{Equilibrium $R_g$ and effective $\tilde{R}_g$ during the translocation process. The 
absolute value of the effective $\tilde{R}_g$ is larger than the equilibrium 
value as it is pulled, but both of them scale with the equilibrium Flory exponent. 
The insets shows the average time 
dependence of the $v_{CM}(t)$ and $R_g(t)$ during the translocation. The straight lines in the 
inset represent  the average value.}  
\label{rgvz}
\end{center}
\end{figure}

In order to verify this issue first, we have calculated the equilibrium $\langle R_g \rangle $ 
of the chain clamped at one end at the pore in presence of the same LJ wall. We find 
$\nu \sim 0.6 \pm 0.01$ (Fig.~\ref{rgvz}).  We have also calculated the 
relaxation time $\tau_r$ of the end-to-end vector 
$\rm \langle (\mathbf{R}_{1N}(t+\tau) - \langle \mathbf{R}\rangle) \cdot 
(\mathbf{R}_{1N}(t+\tau) - \langle \mathbf{R}\rangle) \rangle \sim \exp(-t/\tau_r)$ and checked 
that we get the same $\nu$ from the relaxation measurements. 
This is consistent with the theoretical prediction of Eisenriegler, Kremer, and Binder that in 
presence of the wall the exponent $\nu$ remains the same as that of its bulk counterpart
\cite{Eisenreigler}. To get an idea 
how fast is the translocation process, compared to the corresponding relaxation time, 
for the chain lengths $N= 64, 128$, and $256$, 
we find $\tau_r \sim  1000, 4500$, and $20200$ respectively 
and the corresponding average translocation times $\langle \tau \rangle$ are $215, 530$, and $1330$,
respectively. Even in the linear response regime where $\langle \tau \rangle \sim 1/F$, 
we observe $\tau_r >> \langle \tau \rangle$.  
The insets of Fig.~\ref{rgvz}
shows the time dependence of  $\tilde{R}_g(t)$ (we use 
a different notation $\tilde{R}_g$ for the driven chain). We notice  
that during the translocation process
the chain is significantly elongated around $t \simeq 0.5\langle \tau \rangle$ and  
acquires relatively compact structure immediately 
upon exiting the pore. The dashed lines (black and green) show 
the corresponding average values $\tilde R_g$ from which we extract the 
exponent $\bar\nu \simeq \nu$ (Fig. ~\ref{rgvz}). 
Contrary to what is assumed by 
Kantor and Kardar, we notice 
significant distortion of the chain. Surprisingly,   
we find that the $\tilde{R}_g$ scales almost 
the same way as $\langle R_g \rangle \sim N^{0.6}$ (slopes are the same in Fig. 4).  
even when $ \langle \tau \rangle << \tau_r$. Therefore, numerically we find that the chain is 
still described by the equilibrium $\langle R_g \rangle$. 

Likewise, as expected in LD simulation, we notice that $v_{CM}(t)$ saturates quite quickly 
and this value is almost the 
same during the translocation process and $ \simeq \langle v_{CM} \rangle \sim 1/N^\delta $. 
where $\delta = 0.81 \pm 0.04$. Since 
$\bar{\nu} \simeq \nu$, our studies indicate that it is the $ \langle v_{CM} \rangle $ 
which does not exhibit 
inverse linear dependence on chain length $N$ is the responsible factor 
for the deviation from $\langle \tau \rangle \sim N^{1+\nu}$.   

It is worth mentioning that we have carried out exactly the same LD simulations with 
wall particles on a square lattice\cite{Aniket}.    
We find that 
$\langle \tau \sim N^{1.41}$,   
$\langle s^2(\tau)\rangle \sim \tau^{1.52}$, and $\langle \Delta s^2(\tau)\rangle \sim \tau^{1.45} $
(so that $\alpha\beta \simeq 2.0$, same as reported here if we extract $\beta$ from 
the slope of the plot $\langle \Delta s^2(\tau)\rangle \sim \tau $). 
These results for the square pore are also consistent with recently reported LD and 
MD simulation results in 3D using GROMACS\cite{Kaifu0805}.
Recently Gauthier \textit{et al.}\cite{Slater} carried out similar studies of polymer translocation 
through a narrow pore (including hydrodynamic 
interactions) and found a systematic variation of the measured scaling exponents as a 
function of the pore width. However, their studies are limited to relatively narrow range of 
$N$ up to 31 only. In our studies the exponents for a relatively wide range of $N$  
seem to depend on the pore geometry. Whether this implies true nonuniversality or not remains an open 
issue. 
\section{CONCLUSION} 
To summarize, we have used Langevin dynamics in 3D to study the scaling properties of a driven 
translocating chain through a nanopore. 
We notice that the chain undergoes a significant shape 
change during the fast translocation process, contrary to what assumed by Kantor and Kardar 
is formulating 
the theory of forced translocation. However, despite significant distortion, we find the chain  
is still described by the equilibrium Flory exponent. 
We find that the $\langle v_{CM} \rangle$ does not scale as its bulk counterpart and 
depends on pore width and geometry.  
It is likely that density variation on either side of the pore 
during the translocation process affects the overall motion of the chain. 
We find that the $\alpha = 1.36 < 1+\nu$. 
It is worth mentioning that the collective numerical work by various groups failed to validate 
the Kantor and Kardar result $\alpha = 1+\nu$ for the forced translocation, including the results 
listed here.  Likewise,   
although the value of $\alpha=1.36$ that we obtain is in excellent agreement with the 
analytical estimate of Vocks \textit{et al.} $\alpha = \frac{1+2\nu}{1+\nu} = 1.37$ in 3D, 
the results from 2D simulations do not agree with the estimate of Vocks \textit{et al.}.    
Finally, we notice a difference in the $s$-exponent $\beta$ when calculated from its second moment 
($\beta = 1.6$) and its fluctuations ($\bar\beta = 1.44$). The later ($\bar\beta = 1.44$) 
agrees with $\alpha\bar\beta \simeq 2.0$ while $\beta = 1.6$ overestimates it 
($\alpha\beta \simeq 2.2 > 2.0$). The fluctuations $\langle \Delta s^2 (\tau) \rangle$ 
seem to reveal more structures not adequately studied so far.  
When we compare these results with the existing theories 
and other numerical results we notice that 
these results only partially support one theory or the other. 
Certainly more numerical and analytic 
work are needed for a more comprehensive understanding of forced translocation through nanopore. \par
\section{ACKNOWLEDGEMENT}
A. B. gratefully acknowledges the local hospitality 
of the Institut f\"{u}r Physik, Johannes-Gutenberg Universit\"{a}t, Mainz, the 
travel support from the Deutsche Forschungsgemeinschaft, SFB 625/A3, and 
the local hospitality and travel support from the COMP Center of Excellence, Helsinki University of Technology
respectively, and thanks Prof. M. Muthukumar for valuable discussions.       
T.A-N. and K.L. have been in part supported by the Academy of Finland through the 
COMP Center of Excellence program and TransPoly consortium grant.

\end{document}